\documentstyle[prl,twocolumn,aps,epsf]{revtex}
\def\break#1{\pagebreak \vspace*{#1}}
\begin{document}
\draft
\title{Effective low-energy theory for correlated carbon nanotubes}
\author{Reinhold Egger$^1$ and Alexander O. Gogolin$^2$}
\address{${}^1$Fakult\"at f\"ur Physik, Albert-Ludwigs-Universit\"at, 
Hermann-Herder-Stra{\ss}e 3, D-79104 Freiburg, Germany\\
${}^2$Department of Mathematics, Imperial College, 180 Queen's Gate,
London SW7 2BZ, United Kingdom}
\maketitle
\widetext
\begin{abstract}
The low-energy theory for 
single-wall carbon nanotubes including Coulomb
interactions is derived and analyzed.
It describes two fermion chains without interchain hopping but
coupled in a specific way by the interaction.
The strong-coupling properties are studied by
bosonization, and consequences for
experiments on  single armchair nanotubes are discussed.
\end{abstract}

\narrowtext

The recent discovery of carbon nanotubes \cite{iijima}
has sparked a tremendous amount of activity.
Nanotubes are nanoscale particles obtained by wrapping
a single layer of graphite into a cylinder.  The
electronic properties of a $(n,m)$ tube are determined by the
integer indices $0\leq m \leq n$ of the wrapping
superlattice vector, and
depending on the choice of $m$ and $n$, the tube
is either a metal, a narrow-gap semiconductor,
or an insulator \cite{hamada}.
Experimentally, nanotubes can be produced
using  the carbon-arc technique \cite{iijima}
 or by laser ablation  \cite{thess}
 of Co- or Ni-doped graphite targets. The latter method yields
metallic $(n,n)$ ``armchair'' nanotubes with $n=10$
in rather large quantities, usually deposited in triangular-packed ropes.
Remarkably, the first transport measurements 
on a single 3$\mu$m long $(10,10)$ nanotube have been reported
recently \cite{tans}.

Carbon nanotubes are perfect experimental realizations of one-dimensional 
(1D) conductors. Interacting 1D electrons usually exhibit
Luttinger liquid rather than Fermi liquid behavior
characterized by, e.g., the absence of Landau quasi-particles,
spin-charge separation, suppression of the
electron tunneling density of states, and 
interaction-dependent power laws for transport quantities \cite{book}.
So far non-Fermi liquid behavior has been masked by 
charging effects due to large contact
resistances between the nanotube and the attached leads \cite{tans}.
Nevertheless, future experiments are expected to reveal 
the anomalous conductance laws and related phenomena discussed here
at higher temperature, suitable gate voltages, or smaller
contact resistances. 

The low-energy theory for uncorrelated nanotubes has been
given by Kane and Mele \cite{kane}.  Here we extend
their approach and incorporate Coulomb interactions among the
electrons, focusing on 
$(n,n)$ tubes where interaction effects are most pronounced.
Previously, this problem has only been studied by the perturbative
renormalization group (RG) using a weak short-range (Hubbard) 
interaction \cite{krotov}. 
In this Letter, we discuss arbitrary
interaction potentials and, in particular, the strong-coupling regime
which emerges at low temperatures.
This is of importance as the experiments described
\break{1.0in}
in Ref.\cite{tans} are characterized by long-ranged (unscreened)
Coulomb interactions.
The low-energy theory found here is equivalent to
two  spin-$\frac12$ fermion chains coupled in a rather special
way by the interactions but without interchain single-particle hopping.
In that respect, our theory differs from 
the standard two-chain problem  \cite{fabrizio,schulz}.

\begin{figure}
\epsfysize=6cm
\epsffile{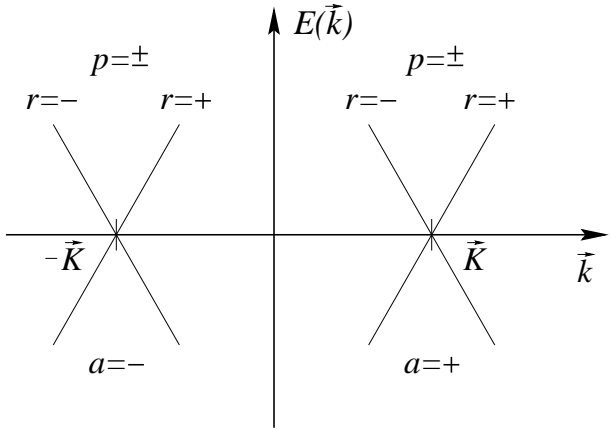}
\caption[]{\label{fig1} 
Low-energy bandstructure of graphite. Fermi points 
are labeled by $\alpha=\pm$, and sublattices 
$p=\pm$ combine to build right- and left-movers ($r=\pm$).}
\end{figure}

The remarkable electronic properties of carbon nanotubes are
due to the special bandstructure of the $\pi$ electrons in
graphite \cite{wallace,divincenzo}. 
 The Fermi surface consists of two distinct Fermi 
points $\alpha {\vec K}$ with  ${\vec K}= (4\pi/3a,0)$
and $\alpha=\pm$, see Fig.~\ref{fig1}.
 Here the $x$-axis points along the tube direction and 
the circumferential variable is $0\leq y \leq 2\pi R$, where
$R=\sqrt{3} na/2\pi$ is the tube radius. The lattice constant is
$a=2.46${\AA}. Since the basis of graphite contains two carbon atoms,
there are two sublattices $p=\pm$ shifted by the vector
${\vec d}=(0,d)$ [where $d=a/\sqrt{3}$], and
hence two degenerate Bloch states $\varphi_{p\alpha}(x,y)$
 at each Fermi point $\alpha\vec{K}$. We follow Ref.\cite{divincenzo}
 and choose these states separately on each
sublattice such that they vanish on the other (note that $K_y=0$),
\[
\varphi_{p\alpha}(x,y) \; \varphi_{-p\alpha'}(x,y) = 0 \;,\quad
\varphi_{p\alpha}(x,y) = \frac{\exp(-i\alpha K_x x)}{\sqrt{2\pi R}}\;.
\]
At low energy scales,  the electron operator
for spin $\sigma=\pm$ can be expanded
in terms of the Bloch waves \cite{kane,divincenzo},
\begin{equation}\label{expa}
\Psi_\sigma(x,y) = \sum_{p\alpha} \varphi_{p\alpha}(x,y) 
\,\psi_{p\alpha\sigma} (x) \;.
\end{equation} 
Quantization of transverse motion gives the wavefunction $\chi(y)
= \exp(i M y)$  with the ``mass'' $M$ specified in Ref.\cite{kane}.
Therefore $\chi=1$ for armchair tubes ($M=0$), see Eq.~(\ref{expa}).
 Excitation of other
transversal bands costs the energy $\approx 10$ eV$/n$, and hence
 a 1D situation arises. The conditions for the
low-energy regime are met even at room temperature 
for $(10,10)$ nanotubes.  
Neglecting Coulomb interactions, the Hamiltonian is \cite{kane,divincenzo}
\begin{equation} \label{h0}
H_0= -v \sum_{p\alpha\sigma} p \int dx \;\psi_{p\alpha\sigma}^\dagger
\partial_x \psi^{}_{-p\alpha\sigma} \;,
\end{equation}
where $v\simeq 6\times 10^6$ m/sec is the Fermi velocity.
Switching from the sublattice
($p=\pm$) description to the right- and left-movers ($r=\pm$) 
indicated in Fig.~\ref{fig1} implies a massless 1D Dirac Hamiltonian.
 
Let us now examine Coulomb interactions mediated by a 
(possibly screened)  potential $U(x-x',y-y')$.
Electrons trapped in nonpropagating
orbitals can be incorporated in terms of a dielectric constant $\kappa$,
but free charge carriers in nearby gates could lead to
an effectively short-ranged potential. 
For the experiments of Ref.\cite{tans}, one has an unscreened
Coulomb interaction, and the dielectric constant can be
estimated as $\kappa\approx 1.4$ \cite{fo1}.  
From Eq.~(\ref{expa}) the interaction contribution reads
\begin{eqnarray}\label{general}
H_I&= &\frac12 \sum_{pp'\sigma\sigma'} \sum_{\alpha_1\alpha_2\alpha_3\alpha_4}
 \int dx dx'\; V^{pp'}_{\{\alpha_i\}}(x-x') \\ \nonumber
&\times& \psi^\dagger_{p\alpha_1\sigma}(x) \psi^\dagger_{p'\alpha_2\sigma'}
(x') \psi^{}_{p'\alpha_3\sigma'}(x') \psi^{}_{p\alpha_4\sigma}(x) 
\end{eqnarray} 
with the 1D interaction potentials
\begin{eqnarray}\label{intpot}
&& V^{pp'}_{\{\alpha_i\}}(x-x') = \int_0^{2\pi R} dy dy' \; 
\varphi^{\ast}_{p\alpha_1}(x,y)
\varphi^{\ast}_{p'\alpha_2}(x',y') \\ &\times& \nonumber
U(x-x',y-y'+p d \delta_{p,-p'} )\; 
\varphi^{}_{p'\alpha_3}(x',y') \varphi^{}_{p\alpha_4}(x,y) \;.
\end{eqnarray} 
Here  inter-sublattice 
interactions involve the shift vector $\vec{d}=(0,d)$.
It is natural to distinguish three processes associated with the
Fermi points $\alpha=\pm$. First, we have ``forward scattering''
($\alpha$FS) where $\alpha_1=\alpha_4$ and $\alpha_2=\alpha_3$.
Second, we have ``backscattering'' ($\alpha$BS) with $\alpha_1=-\alpha_2=
\alpha_3=-\alpha_4$. Finally, at half-filling there is an additional
``Umklapp'' process ($\alpha$US) characterized by 
$\alpha_1=\alpha_2=-\alpha_3=-\alpha_4$.
These processes are different from the conventional 
ones \cite{book} since they do not necessarily mix right- and left-moving 
branches but rather involve different Fermi points, see Fig.~\ref{fig1}.

Let us start with $\alpha$FS. We first define 
\begin{equation}\label{v0}
V_0(x)= \int_0^{2\pi R} \frac{dy}{2\pi R}
\int_0^{2\pi R} \frac{dy'}{2\pi R}\;  U(x,y-y')\; ,
\end{equation}
whence from Eq.~(\ref{intpot}) the $\{\alpha_i\}$-independent
 1D potential reads
$V^{pp'}_{\alpha{\rm FS}}(x) = V_0(x)+\delta_{p,-p'} \delta V_p(x)$
with
\[
 \delta V_p  =  \int_0^{2\pi R} \frac{dy dy'}{(2\pi R)^2}
 [U(x,y-y'+p d)- U(x,y-y') ] \;.
\]
The correction $\delta V_p$
 measures the difference between intra- and inter-sublattice
interactions.
Because of the periodicity of the $y$-integrals, expanding in
powers of $d$ implies $\delta V_p(x)=0$. 
Since the potential $V_0$ does not discriminate among
sublattices, the resulting $\alpha$FS interaction couples only the
total 1D charge densities,
\begin{equation}\label{fs0}
H_{\alpha{\rm FS}}^{(0)} = \frac12 \int dx dx' \, 
\rho(x) V_0(x-x') \rho(x') \;,
\end{equation}
with $\rho(x) = \sum_{p\alpha\sigma} \psi^\dagger_{p\alpha\sigma}
\psi^{}_{p\alpha\sigma}$. 
For an unscreened Coulomb interaction ($a_0\simeq a$),
\begin{equation}\label{unscr}
U(x,y) =\frac{e^2}{\kappa\sqrt{a_0^2+x^2 + 4R^2 \sin^2(y/2R)}}\;,
\end{equation}
the potential (\ref{v0}) becomes 
\[
V_0(x)=\frac{2e^2} {\kappa\pi\sqrt{a_0^2+x^2+4R^2}} 
\,K \left(\frac{2R}{\sqrt{a_0^2+x^2+4R^2}}\right)
\]
with the complete elliptic integral of the first kind $K(z)$.
The Fourier transform $V_0(k)$  for $|kR|\ll 1$ is reminiscent of a 
1D quantum wire \cite{book},
\begin{equation}\label{un2}
V_0(k) =  (e^2/\kappa)\; [ 2|\ln(kR)| + \pi\ln2 ]  \;.
\end{equation}
For $|x|\gg a$, the above continuum argument leading to
 $\delta V_p(x)=0$ safely applies. However, 
for $|x|\leq a$, an additional $\alpha$FS term beyond Eq.~(\ref{fs0}) arises
due to the hard core of the Coulomb interaction,
\begin{equation}\label{fs1}
H^{(1)}_{\alpha{\rm FS}}= - f \int dx\sum_{p\alpha\alpha'\sigma\sigma'}
\psi^\dagger_{p\alpha\sigma}\psi^\dagger_{-p\alpha'\sigma'}
\psi^{}_{-p\alpha'\sigma'} \psi^{}_{p\alpha\sigma} 
\end{equation}
with $f/a = \gamma e^2/R$.
Evaluating $\delta V_p(x=0)$ on the wrapped graphite lattice
using  Eq.~(\ref{unscr}) yields 
\begin{equation}\label{gam}
\gamma= \frac{\sqrt{3} a}{2\pi\kappa a_0} \left[ 1- \frac{1}
{\sqrt{1+ a^2/3 a_0^2}}
\right] \approx 0.1 \;.
\end{equation}
In the language of a Hubbard model, we have $f/a= U-V$ where $U= e^2/R$ 
is the on-site and $V$ the nearest-neighbor Coulomb interaction. 
According to Eq.~(\ref{gam}), this difference is small compared to $U$.

Next we discuss $\alpha$BS contributions.
Since Eqs.~(\ref{general}) and (\ref{intpot}) involve a rapidly 
oscillating factor $\exp[2iK_x (x-x')]$, 
these are local processes which do not resolve sublattices,
\begin{equation} \label{bs}
H_{\alpha{\rm BS}}= \frac{b}{2} \int dx\sum_{pp'\alpha\sigma\sigma'}
\psi^\dagger_{p\alpha\sigma}\psi^\dagger_{p'-\alpha\sigma'}
\psi^{}_{p'\alpha\sigma'} \psi^{}_{p-\alpha\sigma} \;.
\end{equation}
Estimating the coupling constant $b$ from Eq.~(\ref{intpot})
for the unscreened interaction (\ref{unscr}) 
results in $b\approx f$, while for well-screened short-ranged interactions, 
$b\gg f$. Experimentally, by using additional gates, one can easily tune
away from half-filling \cite{tans}. Therefore we 
disregard all Umklapp scattering effects in this paper.

To study this effective low-energy model, 
we employ standard Abelian bosonization \cite{book,schulz}. 
For that purpose we first switch to right- and left-movers
($r=\pm$),
\[
\psi_{p\alpha\sigma} = \sum_r \widetilde{U}_{pr} \psi_{r\alpha\sigma} \;
\quad {\rm where} \quad \widetilde{U}^\dagger 
\sigma_y \widetilde{U} = \sigma_z \;,
\]
which then allow for straightforward bosonization, 
\begin{eqnarray}\label{boson} &&
\psi_{r\alpha\sigma}(x)=\frac{\eta_{r\alpha\sigma}}{\sqrt{2\pi a}}
\exp \Bigl[ i q_F r x + i \frac{\sqrt{\pi}}{2} 
\Bigl(\phi_{c+} + r\theta_{c+}
\\ \nonumber &+&
  \alpha\phi_{c-} +
 r\alpha\theta_{c-} + \sigma \phi_{s+} + r\sigma \theta_{s+}
+ \alpha\sigma \phi_{s-} + r\alpha\sigma\theta_{s-} \Bigr) \Bigr]\;.
\end{eqnarray}
Phase fields  for  the total and relative 
($\delta=\pm$) charge ($j=c$) and
spin ($j=s$) channels obey the algebra
\[
[\phi_{j\delta}(x),\theta_{j'\delta'}(x')] = -(i/2) \delta_{jj'}\delta_{
\delta\delta'}\, {\rm sgn}(x-x') \;.
\]
Thus $\phi_{j\delta}$ and $\theta_{j\delta}$ are dual fields.
Density and current operators can be written in terms of the
$c+$ field,
\[
\rho(x)=(2/\sqrt{\pi})  \partial_x \theta_{c+}(x) \;, \qquad 
I=(2e/\sqrt{\pi}) \partial_t \theta_{c+}(0) \;.
\]
The wavevector $q_F$ is related to deviations $\delta \rho=4q_F/\pi$
of the average electron density from half-filling and
can be tuned by gates. Finally, the $\eta_{r\alpha\sigma}$
are Majorana fermions ensuring anticommutation relations
among different $r\alpha\sigma$. Since the Hamiltonian
contains only the combinations $A_{\pm\pm} = \eta_{r\alpha\sigma} 
\eta_{\pm r\pm \alpha\sigma}$, these 
can be represented using standard Pauli matrices.
Besides $A_{++}= 1$, we have
$A_{+-}=i\alpha\sigma_x, \;A_{-+}=ir\alpha\sigma_z$ and
$A_{--}=-i r\sigma_y$.
Using Eq.~(\ref{boson}), the bosonized expressions then read
\begin{eqnarray} \label{bh0}
 H_0 &=& \frac{v}{2} \int dx \sum_{j\delta} \left[
 ( \partial_x \phi_{j\delta})^2
+ K^{-2}_{j\delta} (\partial_x \theta_{j\delta})^2 \right]\\
\label{bfs0}
H_{\alpha{\rm FS}}^{(0)} &=& \frac{2}{\pi} \int dx dx' \;
\partial_x\theta_{c+}(x) V_0(x-x') \partial_{x'} \theta_{c+}(x')
 \\ \nonumber
 H_{\alpha{\rm FS}}^{(1)} &=& \frac{f}{(\pi a)^2} \int dx \; [
-\cos(\sqrt{4\pi} \, \theta_{c-} ) \cos(\sqrt{4\pi} \, \theta_{s-} ) \\ 
\label{bfs1}
&-&\cos(\sqrt{4\pi} \, \theta_{c-} ) \cos(\sqrt{4\pi} \, \theta_{s+} ) \\
\nonumber
&+&\cos(\sqrt{4\pi} \, \theta_{s+} ) \cos(\sqrt{4\pi} \, \theta_{s-} )]\\
\nonumber
 H_{\alpha{\rm BS}} &=& \frac{b}{(\pi a)^2} \int dx \, [
\cos(\sqrt{4\pi} \, \theta_{c-} ) \cos(\sqrt{4\pi} \, \theta_{s-} ) \\
\label{bbs} &+&
\cos(\sqrt{4\pi} \, \theta_{c-} ) \cos(\sqrt{4\pi} \, \phi_{s-} )\\
\nonumber
&+&\cos(\sqrt{4\pi} \, \theta_{s-} ) \cos(\sqrt{4\pi} \, \phi_{s-} )]\;.
\end{eqnarray}
Although bosonization of Eq.~(\ref{h0}) gives $K_{j\delta}=1$,
interactions renormalize these parameters. 
In particular, $H_{\alpha{\rm FS}}^{(0)}$ can be incorporated into $H_0$
by putting $K_{c+}= K = 1/\sqrt{1+4V_0(k\simeq 0)/\pi v} <1$,
while for all other channels, $K_{j\delta}=\sqrt{1+f/\pi v}>1$. 
For the long-ranged interaction (\ref{unscr}), one has
$K = {\rm const.}/\sqrt{|\ln [{\rm max}(T,v/L)]|}$ 
\cite{book} such that at zero temperature, $K$
vanishes in a very long tube. 
Velocity renormalizations are ignored here as they do not 
change exponents.

From the perturbative RG equations \cite{unp}, it follows
that around the Gaussian point $f=b=0$,
 $H_{\alpha{\rm FS}}^{(1)}$  is irrelevant,
while $H_{\alpha{\rm BS}}$  scales to strong coupling, 
$b\to \infty$. Let us first examine short-ranged interactions
such that $f \ll b$.  For $f=0$, the relative charge and spin channels can be 
refermionized in terms of four Majorana fermions $\xi_{j R(L)}(x)$.
Here $j=1,2$
corresponds to $(s-)$, $j=3,4$ to $(c-)$, and
$R(L)$ denotes the right/left-moving part.
Remarkably, refermionization  shows that the last term in Eq.~(\ref{bbs})
vanishes identically, and we are left with
\begin{eqnarray}
 \nonumber
H(c-,s-) &=& - \frac{i}{2} \sum_{j=1}^4 \int dx\, \left( \xi_{jR}\partial_x
\xi_{jR} - \xi_{jL} \partial_x \xi_{jL} \right)  \\
&-& 2 b \int dx\, \left(\xi_{3R}
\xi_{3L} + \xi_{4R}\xi_{4L}\right) \xi_{1R}\xi_{1L}  \;.
\label{maj}
\end{eqnarray}
The Majorana fermion $\xi_2$ stays massless,
i.e., the $(s-)$ sector carries one massive and
one massless branch. This behavior is due to the symmetric 
appearance of the dual fields $\theta_{s-}$ and $\phi_{s-}$ in
Eq.~(\ref{bbs}) which does not permit
complete pinning due to the Heisenberg uncertainty relation \cite{schulz}.
The $(c-)$ sector is fully gapped, and we can put $\theta_{c-}=0$. 
The masses of the three massive Majorana fermions $\xi_{1,3,4}$ at this
strong-coupling point are approximately equal $(\approx m_b)$
and can be estimated from mean-field theory
as $m_b  \approx \omega_c \exp(-\pi v/\sqrt{2} b)$,
where $\omega_c = 7.4$ eV is the bandwidth of the $\pi$ 
electrons \cite{hamada}.
Correlation functions for the $(s-)$ sector follow from
the correspondence between two Majorana fermions and
the order/disorder operators of the 2D Ising model \cite{schulz}.
In particular, one has scaling dimension $\eta=1/8$   
for  $\sin(\sqrt{\pi} \theta_{s-})$ and
$\sin(\sqrt{\pi} \phi_{s-})$ [their scaling dimension for $b=0$ is 
$\eta=1/4$], while correlation functions
of $\cos(\sqrt{\pi} \theta_{s-})$ and $\cos(\sqrt{\pi} \phi_{s-})$
decay exponentially. In the $(c-)$ sector, correlations of 
$\cos(\sqrt{\pi} \theta_{c-})$ show long-range order while all
other operators lead to exponential decay.

While $H_{\alpha{\rm FS}}^{(1)}$ is irrelevant at $f=b=0$, 
it becomes relevant at the new strong-coupling point
which therefore describes only an intermediate fixed point.  
The term $\sim \cos(\sqrt{4\pi} \theta_{s+})
\cos(\sqrt{4\pi} \theta_{s-})$ in Eq.~(\ref{bfs1}) stays marginal,
but the two other terms become relevant with scaling dimension
$\eta=1$. Therefore  $\xi_2$ as well as
the $(s+)$ field  acquire the mass $m_f\approx (f/b) m_b$. 
At the emerging strong-coupling fixed point, we have
 long-range order in the operators $\cos(\sqrt{\pi} \theta_{s+})$,
$\cos(\sqrt{\pi} \theta_{c-})$ and $\sin(\sqrt{\pi} \phi_{s-})$,
with exponential decay in all other operators 
except those of the $(c+)$ sector.

This analysis for a screened interaction potential predicts that
the exponents corresponding to
the first (intermediate) strong-coupling point should be observable on
 temperature scales $m_f <  T < m_b$ with a
crossover to a regime $T < m_f$ dominated by the
true $T=0$ fixed point.
For long-ranged interactions, we have  $m_f\approx m_b$, and
the  intermediate fixed point 
and the associated crossover phenomenon are then absent \cite{fo3}.

With the strong-coupling solution discussed above, we can now
examine temperature-dependent susceptibilities and
other experimentally accessible quantities of interest. 
At temperatures $ T > m_b$, the
dominant correlations come from the inter-sublattice 
charge-density wave (CDW) and spin-density wave (SDW) operators
\[
\hat{O}_{CDW} \sim \sum_{p\alpha \sigma} \psi^\dagger_{p\alpha \sigma}
\psi^{}_{-p \alpha \sigma} \;, 
\hat{O}_{SDW} \sim \sum_{p\alpha \sigma}  \sigma 
\psi^\dagger_{p\alpha \sigma}\psi^{}_{-p \alpha \sigma} \;, 
\]
with $\langle \hat{O}(x) \hat{O}(0) \rangle \sim \cos(2q_F x) x^{-(K+3)/2}$
for both \cite{fo4}. Due to 
a larger prefactor, SDW correlations will be more pronounced.
For $m_f<T<m_b$, this decay is changed into the slower $x^{-(2K+3)/4}$ law. 
Both the CDW and SDW correlations decay exponentially at very low
temperatures $T<m_f$. However, there is also a $4 q_F$ CDW component
effectively coming from $\hat{O}^2_{CDW}$ \cite{schulz}, which leads
to a slow $\cos(4q_F x) x^{-2K}$ decay at $T<m_b$. This is in fact the dominant
instability at temperatures $T<m_b$ and strong correlations, $K<1/2$.
Remarkably, it is not $K_x$ but $q_F$ which determines
the period of all the dominating correlations. Since $a q_F \ll 1$,
one has pronounced {\em ferromagnetic} correlations. This offers
an explanation for the ferromagnetic tendencies observed in Ref.\cite{tans}.

Superconductivity (SC) has been predicted to be
 quite a robust feature of two-chain models \cite{krotov,fabrizio,schulz}.
In the nanotube, the dominant SC correlations come from the
intra-sublattice singlet SC pairing operator
\[
\hat{O}_{SC} \sim \sum_{p\alpha\sigma} \sigma \psi^{}_{p\alpha\sigma} 
\psi^{}_{p-\alpha-\sigma}\;,
\]
 while triplet SC plays no
role. For $T>m_b$, correlations
decay faster than $x^{-2}$, and for $m_f<T<m_b$, their
$x^{-(2/K+3)/4}$ decay is subdominant to the CDW and SDW
correlations.  Finally, at very low temperatures, $T<m_f$, 
we obtain $\langle\hat{O}_{SC}(x)\hat{O}_{SC}(0)\rangle
\sim  x^{-1/2K}$. Therefore SC becomes the dominant instability only
at very low temperatures and for sufficiently short-ranged
interactions ($K>1/2$). For long-ranged interactions, 
superconducting correlations are of no importance.

The properties of the above strong-coupling points also
determine conductance laws. In the
absence of impurities, one has perfect conductance quantization,
$G=G_0= 2e^2/h$, and  experimentally measured resistances are
contact resistances. 
There are nevertheless various sources for impurities,
 e.g.,  structural imperfections,  charge defects in the substrate, 
 topological defects (dislocation pairs)
\cite{iijima}, or  twists \cite{kane}. Since the first two
act like elastic potential scatterers coupling to the density
operator, from Eqs.~(\ref{expa}) and (\ref{boson}) it is apparent that 
such processes will not exert backscattering in the $(c+)$ channel. Hence
they do {\em not}\, affect the conductance.  The latter two effects, 
however,  mix  $r=\pm$ modes  and thus couple to the $(c+)$ mode.
Both lead to mass-term like perturbations \cite{kane} of the generic form
\[
H_{\rm imp} = v\int dx M(x) \sum_{p\alpha \sigma}
 \psi^\dagger_{p\alpha\sigma} \psi^{}_{-p\alpha\sigma}\;.
\]
Assuming a single static pointlike impurity center,
in  order $M^2$  the following temperature dependence
of the corrections $\delta G$ to $G_0$ is found. For $T>m_b$,
we obtain $\delta G\sim T^{(K-1)/2}$, which is turned into
$\delta G\sim T^{(2K-5)/4}$ at $m_f < T < m_b$. 
At even lower temperature, $T<m_f$, from perturbation theory in $M^2$,
one would get $\delta G\sim T^{K-2}$.
For extremely low temperatures, however,
by invoking the usual duality argument \cite{book},
this is turned into $G (T\to 0) \sim T^{-2+4/K}$, i.e., a 
vanishing conductance similar to the Luttinger liquid case.
For long-ranged interactions, this leads to a pseudogap behavior
at very low temperatures.

To conclude, the effective low-energy theory for correlated
single-wall nanotubes has been given.  Our theory
explains the ferromagnetic tendencies observed in recent
experiments and predicts the temperature dependence of
various susceptibilities and the conductance 
in the presence of impurities.
When additional gates are present, an interesting crossover 
related to screened interactions should be observable.

We wish to thank M.H.~Devoret, D.~Esteve, H.~Grabert, and A.A.~Nersesyan
for helpful discussions.
This work has been partly carried out during an 
extended stay of R.~E.~at the Imperial College 
funded by the EPSRC of the United Kingdom.


\begin{references}
\bibitem{fo1} For a $1/|x|$ interaction, the charging energy of a
tube of length $L$ is $E_c = e^2 \ln(L/R)/
(\kappa L)$. The experimental value $E_c = 2.6$ meV \cite{tans} 
for $L=3\mu$m leads to $\kappa = 1.4$.
\bibitem{fo3} A simple order-of-magnitude estimate
for $1/|x|$ interactions puts $m_f\approx m_b$ into the
mK range.  However, for screened interactions,
$m_b$ can be significantly larger.
\bibitem{fo4} If $\hat{O}$ has scaling dimension $\eta$ [i.e.,
$\langle \hat{O}(x) \hat{O}(0)\rangle \sim x^{-2\eta}$], the corresponding
susceptibility  $\chi \sim T^{2\eta-2}$.
\end{references}
\end{document}